\documentclass[%
reprint, pdftex
amsmath,
amssymb,
aps,
pra
showpacs
]{revtex4-1}

\usepackage{graphicx}
\usepackage{dcolumn}
\usepackage{bm}
\usepackage[usenames,dvipsnames,svgnames,table]{xcolor}
\usepackage[normalem]{ulem}
\usepackage{upgreek}
\usepackage{physics}
\usepackage{appendix}

\begin{document}
	
	\preprint{APS/123-QED}
	
	\title{Prospects of a superradiant laser based on a thermal or guided beam of Sr-88}%
	
	\author{Mikkel Tang$^{1,2}$, Stefan A. Sch{\"a}ffer$^2$, and J{\"o}rg H. M{\"u}ller$^1$}%
	\affiliation{$^1$Niels Bohr Institute, University of Copenhagen, Blegdamsvej 17, 2100 Copenhagen, Denmark\\
	$^2$Van der Waals-Zeeman Institute, Institute of Physics, University of Amsterdam, Science Park 904, 1098 XH Amsterdam, The Netherlands}

	\begin{abstract}
	The prospects of superradiant lasing on the 7.5 kHz wide $^1$S$_0$-$^3$P$_1$ transition in $^{88}$Sr is explored by using numerical simulations of two systems based on realistic experimental numbers. One system uses the idea of demonstrating continuous superradiance in a simple, hot atom beam with high flux, and the other system is based on using ultra-cold atoms in a dipole guide. We find that the hot beam system achieves lasing above a flux of $2.5 \cross 10^{12}$ atoms/s. It is capable of outputting hundreds of nW and suppressing cavity noise by a factor of 20-30. The second order Doppler shift causes a shift in the lasing frequency on the order of 500 Hz. For the cold atom beam we account for decoherence and thermal effects when using a repumping scheme for atoms confined in a dipole guide. We find that the output power is on the order of hundreds of pW, however the second order Doppler shift can be neglected, and cavity noise can be suppressed on the order of a factor 50-100. Additionally we show that both systems exhibit local insensitivity to fluctuations in atomic flux.
	\end{abstract}
	
	\pacs{}
	\keywords{}
	\maketitle
	
	\section{Introduction}
	The precision of frequency references based on optical cavities is limited by thermal fluctuations in the mirrors \cite{MateiPRL2017}. This motivates the development of frequency references based on superradiant lasers with reduced sensitivity to cavity noise \cite{HollandLaser}. These operate using atomic transitions which are much more narrow than the cavity linewidth, so that the spectral properties of the atoms dominate over the resonator. Alkaline earth atoms such as strontium are a promising source due to a level structure allowing for efficient laser cooling and the availability of narrow, forbidden transitions. So far pulsed \cite{CalciumDelayStatistics, NorciaClockPulses, NorciaClockFreq, NorciaCrossover, qmetLasing, qmetSpectra, CaPulseFeasibility, RamseySuperradiance} and quasi-continuous superradiant lasing \cite{NorciaCrossover,BohnetQuasi} has been investigated extensively, both experimentally and theoretically. Achieving continuous superradiant lasing experimentally is an ongoing challenge, but theoretical studies have both considered continuous superradiance in ultracold atom systems \cite{HollandLaser, KazakovEnsembles, KazakovColdAtoms, ChenCaLaser, JingbiaoChen2005, Debnath, ZeemanSuperradiance, HotterDipoleInteracting, MeiserSteadyState, HaakeSuperradiantLaser, SuperradiantOptoPhases}, and in hot atom systems \cite{ruggedLaser, ChenActiveClock, JagerContLasing, JagerDetunedLasing, bistable} that may be simpler to realize and with higher possible atom fluxes. Some studies have also specifically investigated the impact of inhomogenous broadening effects \cite{BychekInhomogenous, KazakovStability}. The aim of pushing towards the development of continuous superradiance has also sparked development in sources of atomic beams \cite{Bel21, guidedBeam} which can meet the technical requirements for lasing on very narrow transitions.\\
	
	Here we investigate the prospects of continuous superradiant lasing on the $^1$S$_0$-$^3$P$_1$ intercombination line in $^{88}$Sr. We aim to capture all the vital physical effects present in a realistic physical system \cite{guidedBeam, ShengDAMOP} and to bridge the gap to more idealized theoretical descriptions which often omit e.g. details about phase-space distributions, thermal effects, optical forces and realistic energy-level schemes for the sake of simplicity and computational advantages. We consider a cold atom system based on the source in \cite{guidedBeam}, and a hot atom system \cite{ShengDAMOP} based on a recent theoretical proposal \cite{ruggedLaser}. We use numerical simulations to investigate these two systems, the requirements to overcome the lasing threshold, and the influence of cavity fluctuations. In Section II we present a simple cooperativity model for the cold atom system which highlights the parameter regime that is necessary to obtain lasing. In Section III we introduce the full theoretical model used to study superradiant lasing quantitatively in the cold and hot atom systems. In Section IV and V we present the expected lasing dynamics and relevant physical effects for the cold and hot beam systems, respectively.\\
	
	\section{System and Cooperativity Model}
	The systems we describe here consist of $^{88}$Sr atoms propagating through a cavity while initially in the excited $^3$P$_1$ state. In this section we will consider the cold-atom system. To achieve superradiant lasing the atoms must preferentially emit into the cavity mode, and this process competes with emission into the environment and decoherence. In terms of the collective cooperativity we can write this condition as:
	
	\begin{eqnarray} \label{eq:coop}
	C_0 N \gamma \gg \Gamma_{decoh}.
	\end{eqnarray}
	
	Where $C_0$ is the single-atom cooperativity, N is the number of atoms, $\gamma$ is the decay rate of the excited energy level and $\Gamma_{decoh}$ is the dephasing rate of the dipole, including e.g. Doppler broadening and decoherence from pumping. In terms of $\gamma$, the cavity linewidth $\kappa$ and the atom-cavity coupling $g$, we have $C_0 = 4g^2/\gamma \kappa$. In a realistic scenario the cavity mode is a standing wave and has a Gaussian intensity distribution, giving a spatial dependence $g(r,z)$. We approximate the atomic beam density profile by a Gaussian in the plane perpendicular to its propagation axis, characterized by the standard deviation $\sigma_y$ in the dimension that is also perpendicular to the cavity axis, leading to an overlap integral with the cavity mode. Thus if we have a beam of excited Sr atoms propagating at a speed $v$, we can rewrite the lasing condition in terms of experimental parameters as:
	
	\begin{eqnarray} \label{eq:coop2}
	\frac{6\sqrt{2\pi} c^3}{\omega^2} \frac{\Phi}{L W \kappa v} \sqrt{\frac{\frac{W}{\sigma_y}}{\left(\frac{W^2}{\sigma_y^2}\right)^2 + 4}} \frac{\gamma}{\Gamma_{decoh}} \gg 1.
	\end{eqnarray}
	
	The first fraction is constant for a given transition frequency $\omega$. In the second fraction we see that the cooperativity increases linearly with the flux $\Phi$, and decreases if we increase the mode volume (cavity length $L$ or waist radius $W$), the loss rate $\kappa$, or propagation velocity $v$, which affects the density for a fixed flux. The term in the square root is a scaling between 0 and 1 related to the overlap of the atomic beam with the cavity: If the cavity waist is much larger than the atomic beam it yields 1, but if it is much smaller than the beam it approaches 0. The final term $\gamma/\Gamma_{decoh}$ can approach 1/2 (for the lowest possible decoherence rate), but given a finite temperature T and a repumping rate from the ground state $w$ the term will generally be smaller. If the lasing condition is fulfilled and the atom-cavity overlap is good, the output power from the cavity can approach the limit set by energy conservation of $P_{max} = \hbar \omega \Phi$. If repumping is included at the rate $w$ and we assume only atoms within the cavity waist participate in lasing, this limit increases to approximately $P_{max} = \hbar \omega \Phi (1 + 2wW/v)$.\\
	
	Using a dipole guide enables confinement of atoms down to tens of micrometers, combined with propagation velocities on the order of 10 cm/s and µK temperatures. High atomic flux in such a system has been shown to be possible in \cite{guidedBeam}. Though we considered the cold atom system for now, we can also note the most significant qualitative differences of the hot beam system: in this system the thermal decoherence rate and much higher propagation velocities lead to a much higher flux requirement to achieve lasing. Additionally, the output power when lasing cannot exceed $\hbar \omega \Phi$ due to the lack of repumping. However, this value is still orders of magnitude higher than $P_{max}$ for the cold beam system due to the much higher flux.
	
	\section{Theoretical Model}
	Our model is based on a Tavis-Cummings Hamiltonian, including an array of filter cavities \cite{Debnath, qmetSpectra} for extracting spectral information:
	
	\begin{eqnarray}
		H &=& \hbar \omega_c a^\dagger a + \sum_{j=1}^N   \hbar \omega_{e} \sigma_{ee}^j
		+ \sum_{k=1}^{N_f} \hbar \omega_{f}^{k} f_k^\dagger f_k
		\\\nonumber
		&+& \frac{1}{2} \hbar \eta \left( a e^{ -i \omega_d t } + a^{\dagger} e^{ i \omega_d t } \right)\\\nonumber
		&+& \sum_{j=1}^N \hbar g_c^j(\mathbf{r_j})  \left( \sigma_{ge}^j + \sigma_{eg}^j \right) \left( a + a^\dagger \right)\\\nonumber
		&+& \sum_{k=1}^{N_f} \hbar g_f  \left( a + a^\dagger \right) \left( f_k + f_k^\dagger \right).
	\end{eqnarray}
	
	Here $\omega_i$ are the resonance frequencies of the cavity (i=c), stationary atom (i=e), driving laser (i=d) or $k$'th filter cavity (i=f), $a$ ($f_k$) is the ($k$'th filter) cavity field annihilation operator and $\sigma_{xy}^j$ is the spin operator of the $j$'th atom. N is the number of atoms in the simulation and N$_f$ is the number of filter cavities chosen to obtain a given spectral resolution. The $j$'th atom couples to the cavity at a rate $g_c^j(\mathbf{r_j})$ which depends on its position at a given time. Each filter cavity is coupled to the main cavity with a tiny constant $g_f$, such that physical back-action on the main cavity can be neglected. $\eta$ is the driving rate of a laser that is used to initiate the lasing dynamics at the start of a simulation, and which is subsequently set to 0. This Hamiltonian is used to derive a set of differential equations for the atom and cavity states in first order mean-field theory, which are numerically integrated over time, while the motion of each individual atom is treated classically. The rate of coupling to the environment (spontaneous emission and cavity leakage) is derived via Lindblad operators. This model has been presented in \cite{qmetLasing, qmetSpectra} where it has been compared to experiments and used to describe pulsed lasing dynamics in the mK temperature regime.\\
	
	In contrast to previous applications of the model, we here describe a system where atoms continually enter and exit a cavity. Thus the number of atoms continually fluctuates: based on the atom flux a number of new atoms are periodically introduced, and existing atoms are deleted if they move significantly out of the cavity mode. Furthermore we here assume atoms are pumped incoherently, and we therefore include a classical driving laser to initiate the lasing dynamics - in a real system this occurs due to quantum fluctuations, which are not included in 1st order mean field theory. This drive is turned off when we evaluate the steady-state parameters.
	
	\section{Lasing from a Guided Cold Beam}
	
	\begin{figure}[t!]
		\includegraphics[width=\columnwidth]{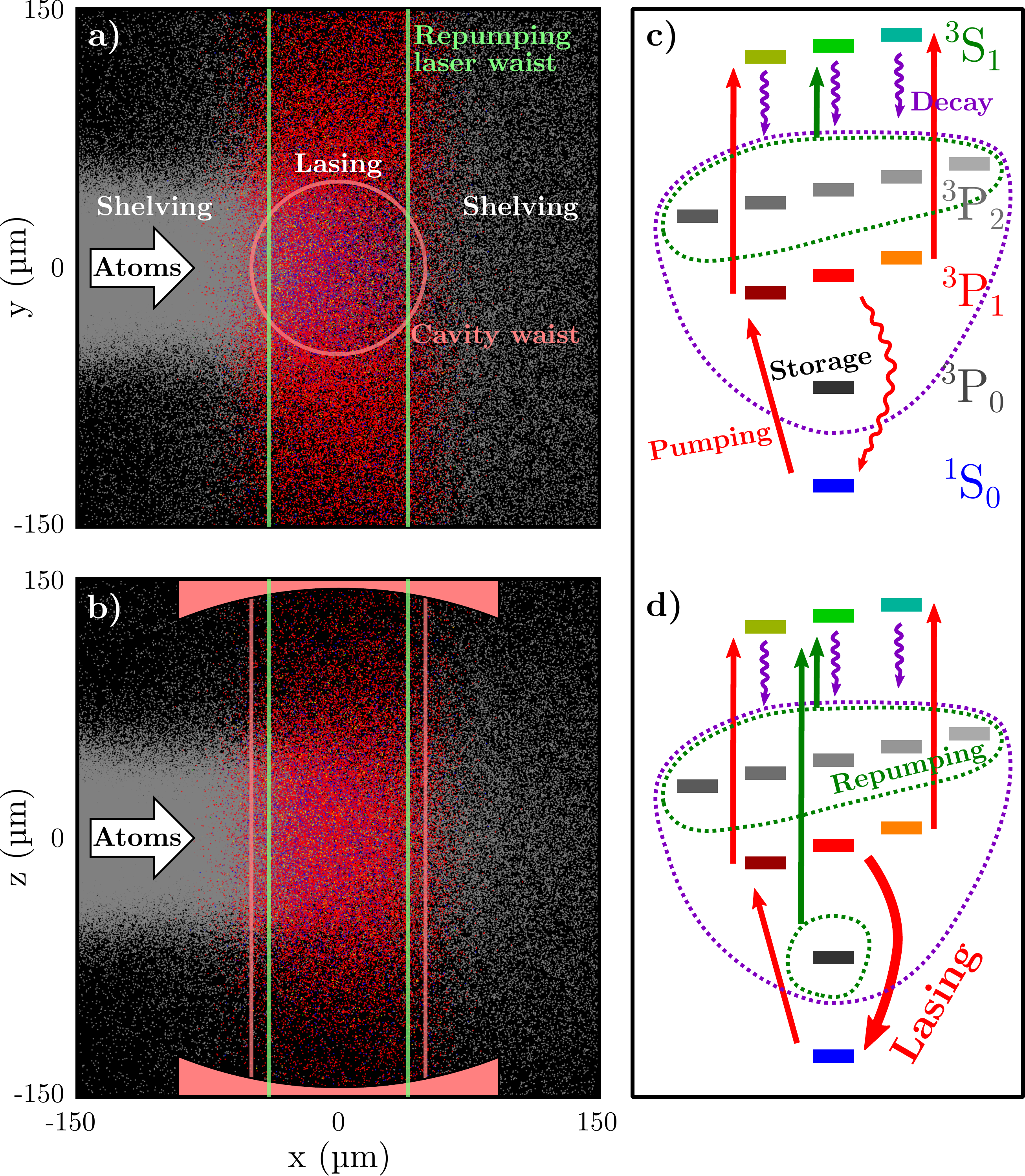}
		\caption{Illustration of the cold beam system, viewed along the cavity axis (a) and the repumping axis (b). Each atom is color coded by its state according to panel (c) and (d) (without the variations in gray), where the shelving and pumping scheme is illustrated. The atoms are initially shelved in $^3P_0$. Within the cavity they are pumped to $^3P_1$ $m_J$=0 and contribute to lasing, heat up and disperse due to repumping - primarily along the repumping axis. (c) Pumping scheme to store atoms in $^3$P$_0$ before entering the cavity. (d) Pumping scheme to transfer atoms to $^3P_1$ $m_J$=0 for the lasing process. \label{fig:guideLatticeBeam}}
	\end{figure}
	
	A flux of up to $\Phi$ = $3 \cross 10^7$ atoms/s has been demonstrated in an guided beam of cold $^{88}$Sr \cite{guidedBeam}, but later improvements to the experimental system have shown that up to $\Phi$ = $3 \cross 10^8$ atoms/s is realistic \cite{Shayne}. We find that to reach the most interesting regime for superradiant lasing it is necessary to continuously repump the atoms in such a system. A repumping scheme which has been used to study quasi-continuous superradiant lasing in a MOT has been demonstrated in \cite{NorciaCrossover}. Since many photon recoils are imparted during each repumping cycle, the atoms can heat up significantly using this scheme. This makes it necessary to shelve the atoms in long-lived states until their position within the Gaussian mode provides a sufficiently high coupling rate for them to emit primarily into the cavity rather than the environment. Based on this we consider the system depicted in Fig. \ref{fig:guideLatticeBeam}. Here the atoms start in the long-lived state $^3$P$_0$ and are continuously repumped within the cavity. To describe this the two-level model is extended to include $^3$P$_0$ and the Zeeman sublevels of $^3$P$_1$, $^3$P$_2$ and $^3$S$_1$, yielding a set of equations for the evolution of atom and cavity states given in the appendix. Each atom is treated separately with its own position, velocity, coupling rate $g_c^j$ to the cavity field, and incoherent driving rates from the repumping lasers. The initial atom velocities are based on a radial temperature of 890 nK and 29 µK along the propagation axis \cite{guidedBeam}, and the distribution in the propagation direction is also treated as Gaussian, with a mean velocity of 0.084 m/s. We include the forces of a dipole guide on the atoms, with a wavelength of 913.9 nm, waist radius of 165 µm and power of 15 W. These forces are calculated on each atom depending on their position and internal state at a given time. Repumping can change the velocity of the initially cold atoms significantly: an average of 21 photon recoils are imparted to bring an atom from $^1$S$_0$ to $^3$P$_1$ $m_J = 0$ (10 from repumping lasers and 11 from decays). The recoils are modeled by using a conditional stochastic master equation (SME) to treat spontaneous decays and the majority of the pumping laser interactions. In this way the model can account for heating of the atoms both due to the recoils, but also due to the abrupt changes in state in the dipole guide potential. To minimize heating along the cavity axis we choose the repumping lasers to be orthogonal to the cavity axis. Collisions between atoms are not taken into account.\\
	
	Two of the pumping laser interactions, from $^1$S$_0$ to $^3$P$_1$ m$_J$=-1 (689 nm) and from $^3$P$_1$ m$_J$=-1 to $^3$S$_1$ m$_J$=-1 (688 nm), are treated using coherent interactions, using the SME quantum jumps only for dissipation. Thus light shifts are included in the model, which can result in shifts in the superradiant lasing frequency. Light shifts on the lasing transition from the 679 and 707 nm repumping lasers can be made negligible ($<$1 Hz) by choosing a suitable "magic" intensity ratio.\\
	
	We assume a magnetic field of 0.476 Gauss pointing along the x axis, giving a 1 MHz splitting of the $^3$P$_1$ Zeeman levels. Thus the dipole guide beam, polarized along the y axis, is magic for the lasing transition. The 688 nm repumping laser is linearly polarized along the x axis, driving the $\pi$ transitions from $^3$P$_1$ (except m$_J$=0 as this is forbidden), while the 689, 679 and 707 nm lasers are polarized along the z axis, driving $\sigma_\pm$ transitions. Since the 688 nm laser has to drive atoms from both m$_J$=-1 and +1 to the corresponding $^3$S$_1$ levels, it is chosen to be equally detuned from both transitions (333 kHz). The other repumping lasers are equivalently tuned symmetrically, except the 689 nm repumper, which is tuned to resonance with the Zeeman-shifted transition of $^1$S$_0$-$^3$P$_1$ m$_J$=-1. Its effect on the m$_J$=+1 transition is neglected due to the large detuning.\\
	
	The performance of the system can be evaluated in terms of the cavity pulling coefficient $c_{pull}$ = $\Delta_L/\Delta_{ce}$, which depends on the shift in lasing frequency, $\Delta_L$, for a given cavity detuning $\Delta_{ce}$. If $c_{pull} = 1$, the cavity fully determines the lasing frequency, and if it is 0, the lasing frequency is determined fully by the atom transition, and the laser is maximally insensitive to cavity length fluctuations. In addition the output power is an important parameter, since a power in the pW regime can be demanding to detect and use as a frequency reference. A third parameter that is experimentally important is the cavity length. A short cavity length gives a stronger atom-cavity coupling, which means the finesse can be lowered to achieve the same lasing threshold, enabling one to operate further in the bad cavity regime. However, a short cavity also requires finer control of the piezo-voltage to keep fluctuations in the resonance frequency acceptable.\\
	
\begin{figure}[t!]
		\includegraphics[width=\columnwidth]{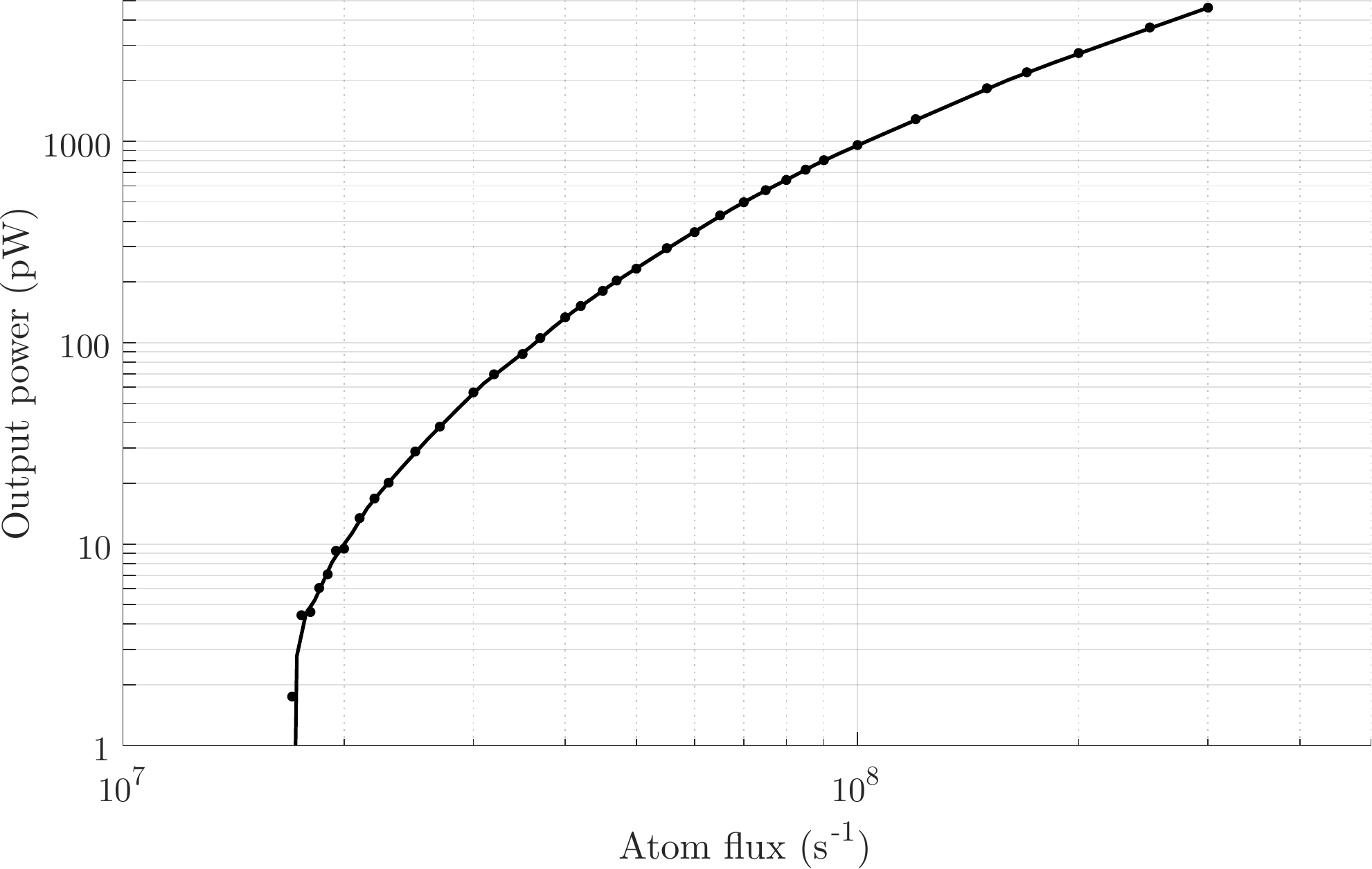}
		\caption{The cavity output power dependency on atom flux $\Phi$. A flux of $10^{8}$ $s^{-1}$ yields $4.6 \cross 10^4$ atoms in the cavity waist in steady state. The line shows a moving mean with a span of two points. \label{fig:coldBeamPout}}
	\end{figure}

    Assuming a cavity length of 25 mm and waist radius of 50 µm we find that the optimal cavity linewidth will be on the order of tens of MHz, and here we choose 20 MHz to obtain a lasing threshold below $2 \cross 10^7$ atoms/s. The expected output power for a given flux is shown in Fig. \ref{fig:coldBeamPout}. Here the 689 and 688 nm pumping lasers have waist radii of 40 µm along the propagation axis and Rabi frequencies $\chi/2\pi$ of 100 kHz and 1 MHz, respectively.\\ 
    
\begin{figure}[t!]
		\includegraphics[width=\columnwidth]{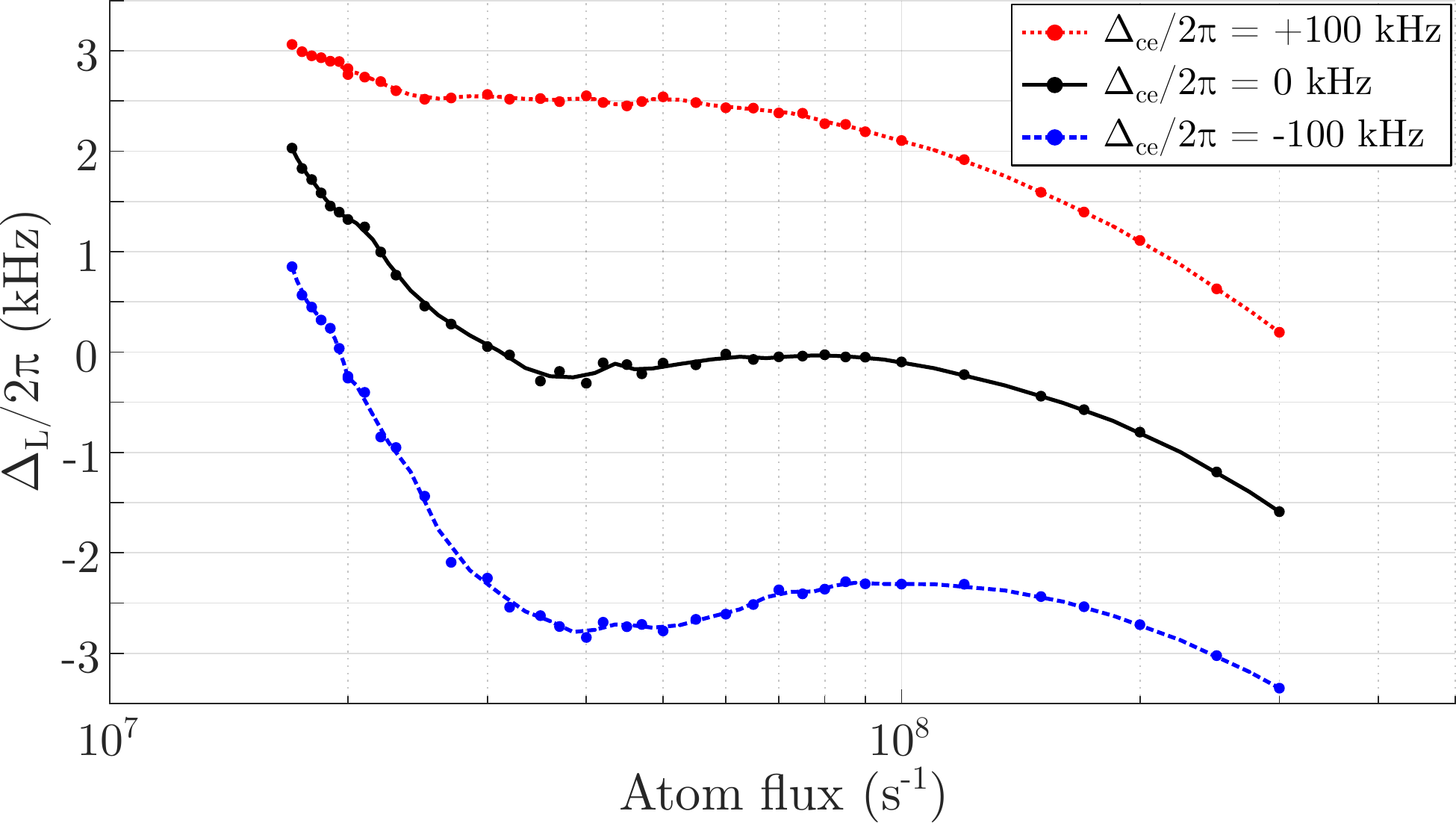}
		\caption{The shift in lasing frequency ($\Delta_L$) for varying atom flux at three different cavity detunings ($\Delta_{ce}$). The local cavity pulling coefficient can be determined by the distance between the curves. Experimental parameters can be chosen to achieve a lasing frequency shift of zero within a range of fluxes, in this case around $8 \cross 10^7$ s$^{-1}$. Lines show a moving mean with a span of two points.\label{fig:coldBeamdetl}}
	\end{figure}

    The lasing frequency obtained in simulations is depicted in Fig. \ref{fig:coldBeamdetl} for varying atom flux at three different cavity detunings. Ideally, the superradiant lasing frequency should have no offset from the atomic transition (corresponding to the middle curve in Fig. \ref{fig:coldBeamdetl} staying at 0). However, we do see offsets which are attributed to light shifts from the pumping lasers. Generally the flux-dependency is nontrivial as the atomic inversion, and thus light shift, depends on the flux. But for a certain choice of repumping rates, the frequency offset is 0 within a range of atom flux, as seen near $8 \cross 10^7$ s$^{-1}$ in Fig. \ref{fig:coldBeamdetl}. Here the lasing frequency shift is also locally independent of the atom flux for small detunings, which removes the sensitivity to noise and slow variations in the flux from the atom source. We find that this local immunity to atom flux fluctuations also depends on a wide number of parameters, such as the sizes of the repumping beams. For smaller repumping regions, the curves in Fig. \ref{fig:coldBeamdetl} become more linear, such that there is no flux-immune plateau when using waists of 30 µm. We attribute these flux-immune properties partially to the flux-dependence of the heating rate from repumping. When the pumping region is smaller, the atoms are heated for a shorter duration, leading to a more monotonic curve for the frequency shifts. Thus the curves can be engineered by variation of experimental parameters.\\
    
    In Fig. \ref{fig:coldBeamdetl} the cavity pulling coefficient can be determined from the relative distance between the curves at different cavity detuning and varies from 0.01 to 0.025. Near the lasing threshold, the atoms do not emit many photons, thus they experience only few repumping cycles and relatively low heating rates. This results in an average temperature within the cavity waist, along the cavity axis, of $T_z=34$ µK for $\Phi = 2\cross 10^7$. As the atom flux increases, the steady state cavity photon number and thus the atomic emission rate $g\sqrt{n}$ increases, leading to more repumping cycles and higher temperatures. The highest cavity pulling coefficients are obtained for a flux near $5 \cross 10^7$ s$^{-1}$, where $T_z = 54$ µK. For even higher flux the cavity pulling starts to decrease again, as the effect of the high number of atoms starts to become more significant than the effects from heating - for a flux of $3 \cross 10^8$ s$^{-1}$, $T_z = 76$ µK.\\
    
    Since thermal effects increase cavity pulling in the system, we additionally investigated the effect of an intra-cavity optical lattice at the magic wavelength of 913.9 nm. Such a lattice could potentially prevent atoms from propagating along the cavity axis significantly. We find that the lattice begins to have a significant impact when the power is in the range of 1-10 W. However we find that the lattice generally increases cavity pulling due to the repumping scheme in combination with the steep optical potentials of the lattice. The fact that the optical potential depth varies for different states leads to significant additional heating, which is revealed by the quantum jumps of the SME.\\
    
    The Purcell rate of $2 \pi \cross$ 82 Hz offers a simple estimate of the linewidth \cite{HollandLaser}, but further narrowing can occur within the superradiant lasing regime \cite{Debnath}, which the cold beam system operates in. The simulations include a number of effects which may also limit the linewidth - the thermal effects and discrete spontaneous emission events from $^3$P$_1$ into the environment. On the other hand they neglect spontaneous emission into the cavity mode. The linewidth in simulations is Fourier-limited to 5 Hz after 200 ms, indicating that the effects that are included do not limit the system beyond the theoretical estimate.\\
    
    \begin{figure}[t!]
		\includegraphics[width=\columnwidth]{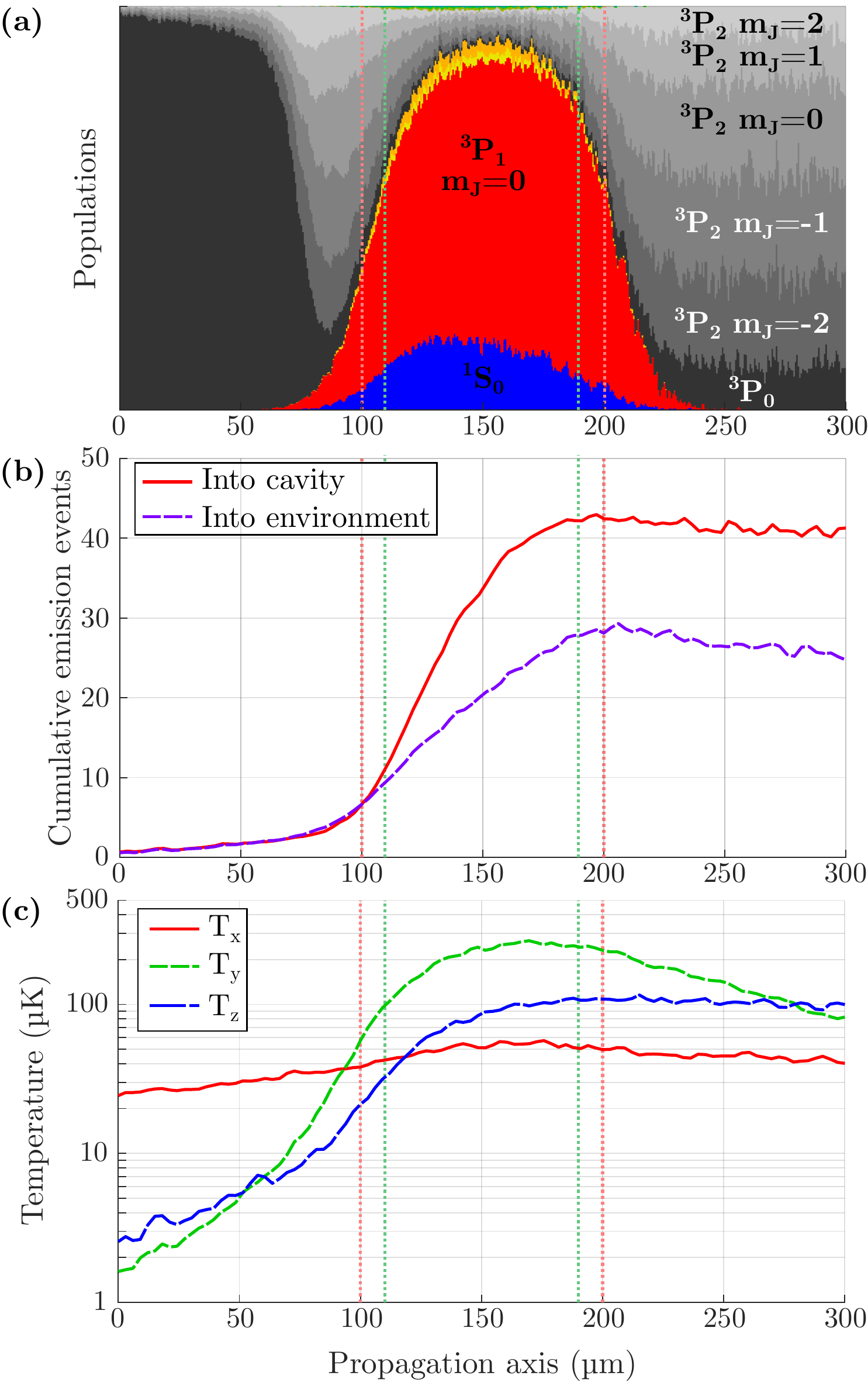}
		\caption{Spatial profile of an actively lasing cold atom beam for a flux of $\Phi$ = $10^8$ atoms/s. (a) Histogram of the spatial distribution of states (yellow is $^3P_1$ $m_J = -1$, orange is $^3P_1$ $m_J = 1$, green are $^3S_1$, others are indicated in the figure). Atoms start in $^1$P$_0$. Once they reach the cavity (waist: red dotted lines), repumping lasers at 679 and 707 nm (waist: green dotted lines) enable inversion on the lasing transition. (b) The mean accumulated number of emission events in the steady-state regime of an atom as function of its x-position, throughout its trajectory up to that point. Note some atoms escape on the left and along the repumping axis near x = 150 µm after interacting with the cavity mode and pumping lasers, while faster atoms tend to interact less and escape on the right. In this regime on the order of 40 photons can be emitted from each atom into the cavity before exiting. (c) Temperature profile along each dimension ($T_x$: propagation axis, $T_y$: repumping axis, $T_z$: cavity axis) as a function of position. The temperature profile is a result of interactions with the repumping lasers and optical dipole guide. \label{fig:spatialGraphs}}
	\end{figure}
    
    The spatial dependence of the dynamics are shown in Fig. \ref{fig:spatialGraphs}, here evaluated for $\Phi = 10^8$ atoms/s. In panel (a) a breakdown of the atomic states is shown as a function of position in the beam. Due to the Gaussian repumping intensity profiles, the $^3$P$_1$ population increases gradually towards the center of the repumping beams. There is additionally a small increase in inversion towards the exit side of the cavity, as the atom beam heats up and interacts less efficiently with the cavity mode. In panel (b) the average number of emission events from $^3$P$_1$ into the cavity (red) or environment (dashed purple) is shown as function of position. For this flux and repumping rate, we see each atom can emit upwards of 40 photons into the cavity and 25 into the environment before escaping. This demands over 1300 photon recoils during the repumping process, highlighting the importance of this heating mechanism. Note that atoms escaping near x=300 µm are generally the fastest and will thus not emit as many photons as those that spend a long time inside the cavity waist, emit many photons, but eventually heat up and escape along the repumping axis. A few atoms also escape from the direction they originally came from and contribute to the nonzero average value for x$<$100 µm. In (c) we show the spatially dependent temperature profile in each dimension. The fact that the atoms are repumped and change state in the guided beam results in non-adiabatic dynamics, because the potential depth is different for each state. Photon recoils from the repumping process mainly heat the atoms along the y axis, but also to a smaller degree along the x and z axes due to the spontaneously emitted photons.
	
	\section{Lasing from a hot atom beam}
\begin{figure}[t!]
		\includegraphics[width=\columnwidth]{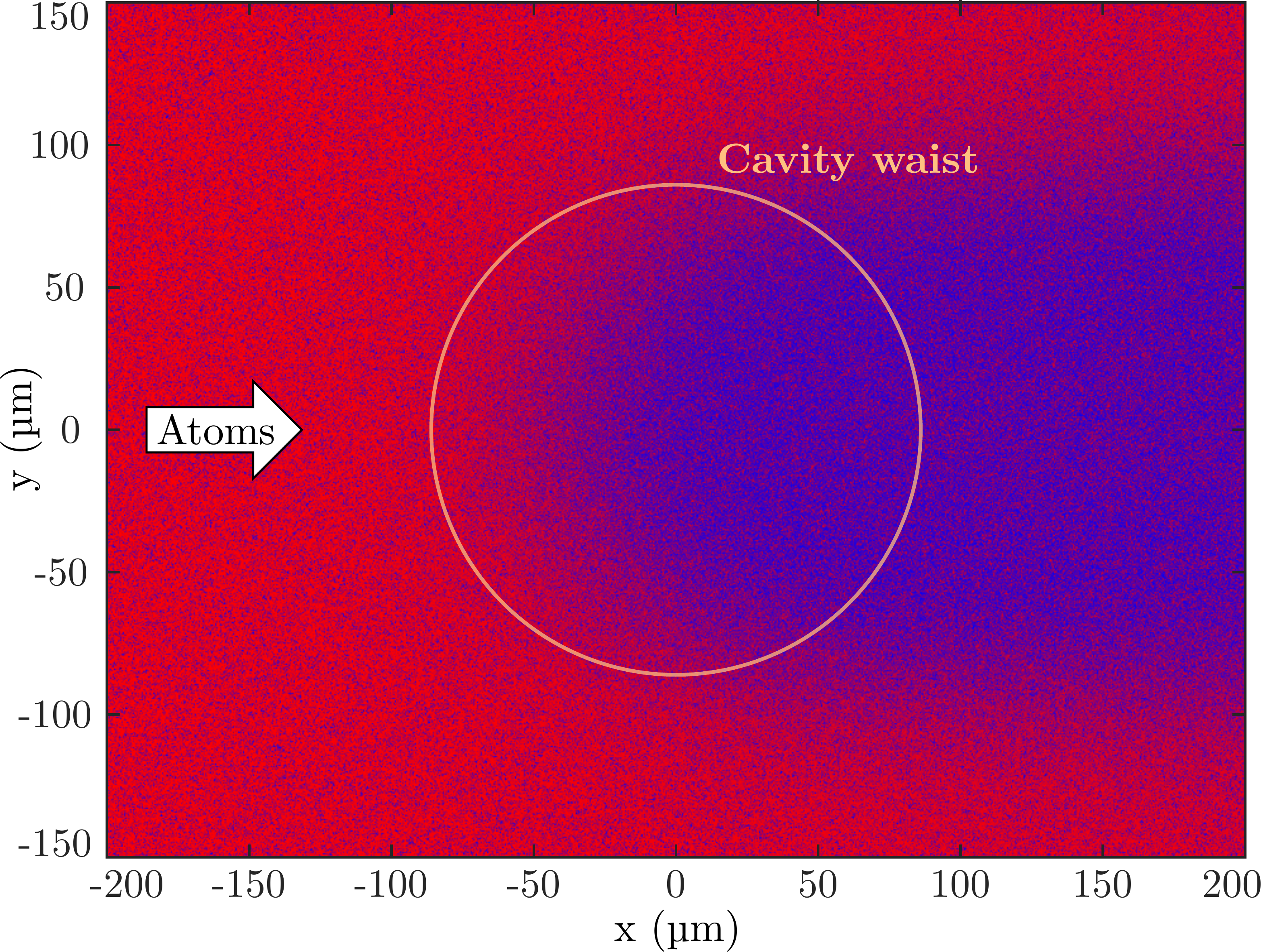}
		\caption{Illustration of the hot beam system. Atoms are color-coded according to their state ($^1$S$_0$ in blue and $^3$P$_1$ in red). Due to velocity selection in this simulation, some of the atoms start in $^1$S$_0$. As the atoms pass through the cavity (waist marked by circle), they will tend to emit a photon and change state to $^1$S$_0$. \label{fig:hotAtomBeam}}
	\end{figure}

	A promising and technically simpler source for superradiant lasing is a thermal beam of strontium atoms emanating directly from an oven, cooled on the $^1$S$_0$-$^1$P$_1$ transition in a 2D molasses to bring the radial temperature to the mK regime. This concept has been explored theoretically in \cite{ruggedLaser, JagerContLasing, JagerDetunedLasing, bistable}. Here we investigate this proposal with the physical constraints of a real system operating on the $^1$S$_0$-$^3$P$_1$ transition of $^{88}$Sr, illustrated in Fig. \ref{fig:hotAtomBeam}. We assume the atoms can start in the excited state $^3$P$_1$ or ground state $^1$S$_0$ and the atoms are treated as a two-level system. Due to the high number of atoms we use a clustering approach for simulating the hot beam system, treating atoms in groups of 100 where each atom in a group has the same position, velocity and internal state. The atom velocities are drawn from Gaussian distributions based on a temperature of 3.6 mK orthogonally to the propagation axis, and from a thermal beam distribution in the propagation direction with a most probable velocity of 400 or 450 m/s. At these velocities the second order Doppler shift becomes significant, and is therefore included in the transition frequency of each atom, typically being on the order of -500 Hz.\\
	
	We can expect that the cavity pulling can be reduced by preventing atoms from moving along the cavity axis. The effect of this motion has been studied for the fast beam, where a bistable regime was found when atoms collectively move across half a wavelength during transit \cite{bistable}. Thus to reduce the influence of atoms that move further than half a wavelength while traversing the cavity waist, we include a velocity selection stage \cite{Bel21} in our simulations. In this scheme the atoms are initially shelved in a long-lived state according to the criterium $\abs{v_z} < \lambda \cross v_{px} / 4 W$, where $v_z$ is the velocity along the cavity axis, $v_{px}$ is the most probable velocity in the propagation direction, and $W$ is the cavity waist radius. The atoms that move too quickly along the cavity axis thus remain in $^1$S$_0$. These atoms are subsequently shifted in momentum space using a resonant laser on the $^1$S$_0$-$^1$P$_1$ transition such that they do not interact with the cavity photons. This requires a Doppler shift significantly greater than the power broadening of the lasing transition due to the intracavity field, for which a push by a few m/s is sufficient in the considered regime. Since this is done by a laser along the cavity (z) axis, the selection is imperfect, as some atoms may move slow enough along z to not be selected, but sufficiently slow along x that they still cross half a wavelength.\\
	
\begin{figure}[t!]
		\includegraphics[width=\columnwidth]{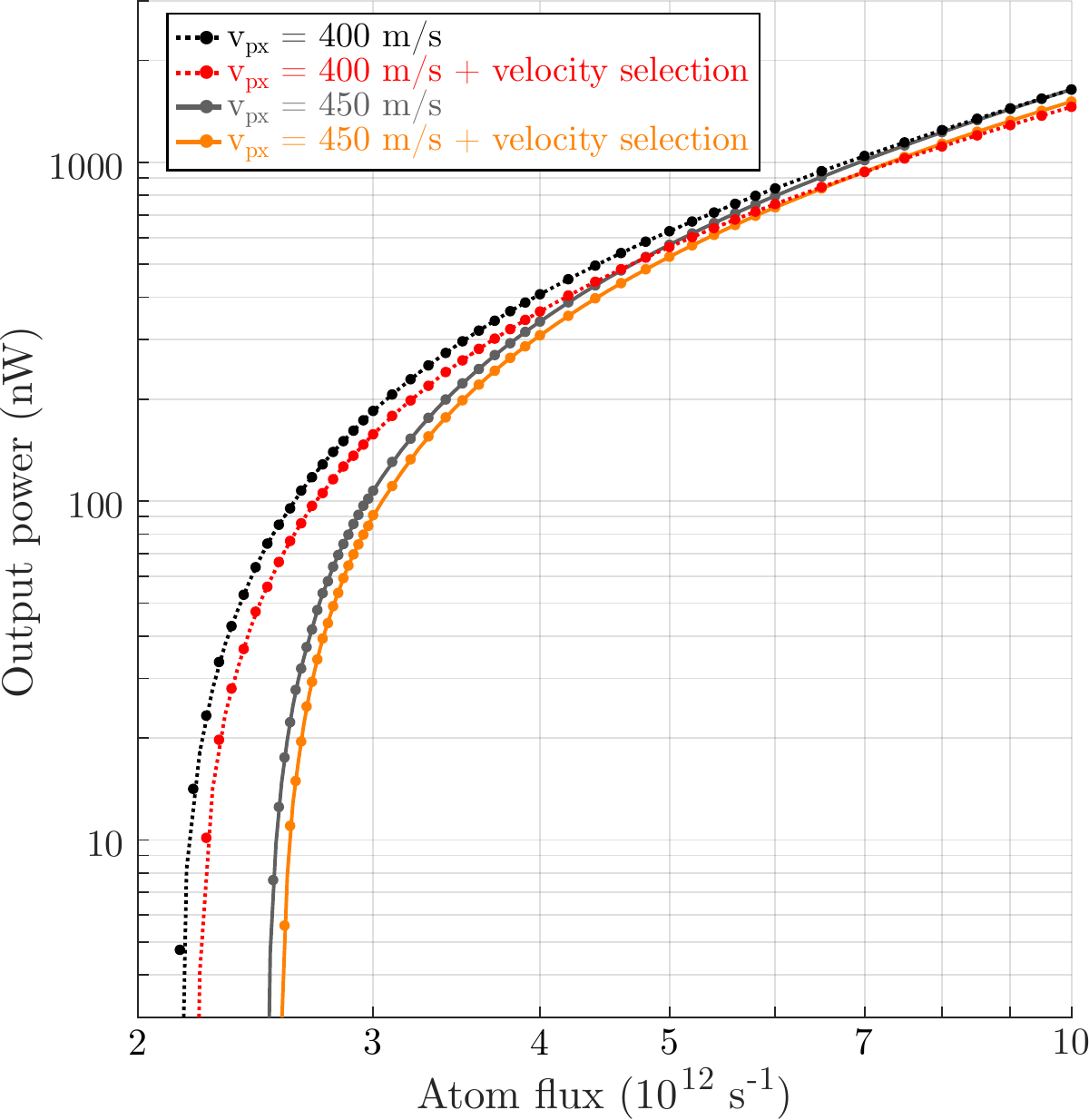}
		\caption{Expected cavity output power as function of atom flux through the cavity mode waist for two values of the most probable atomic velocity, with and without the velocity selection scheme. A flux of $10^{13}$ $s^{-1}$ corresponds to $\left\{3.71,3.28\right\} \times 10^6$ atoms in the cavity waist for $v_{px}= \left\{400,450\right\}$ m/s. The velocity selection scheme leads to a decrease in output power and slight increase in lasing threshold. Lines show a moving mean.\label{fig:hotBeamPout}}
	\end{figure}

	Since multiple stages of loading atoms into magneto-optical traps and guided beams are not necessary, a much higher atomic flux can be obtained through the cavity than in the cold beam system, thus expected values are on the order of $10^{12}-10^{13}$ atoms/s \cite{Bel21}. For this system, with a cavity length of 27.36 mm, linewidth of 2$\pi \cross$53.9 MHz and waist radius $W$ = 86 µm, we find the lasing threshold is reached between 8 and 8.5 $\cross 10^{5}$ atoms within the cavity waist. This atom number is reached for different atom flux depending on the mean propagation velocity (see Fig. \ref{fig:hotBeamPout}). The threshold flux is on the order of $\Phi = 2.5 \cross 10^{12}$ atoms/s and an output power on the order of 1.5 µW can be achieved at $10^{13}$ atoms/s.\\
	
\begin{figure}[t!]
		\includegraphics[width=\columnwidth]{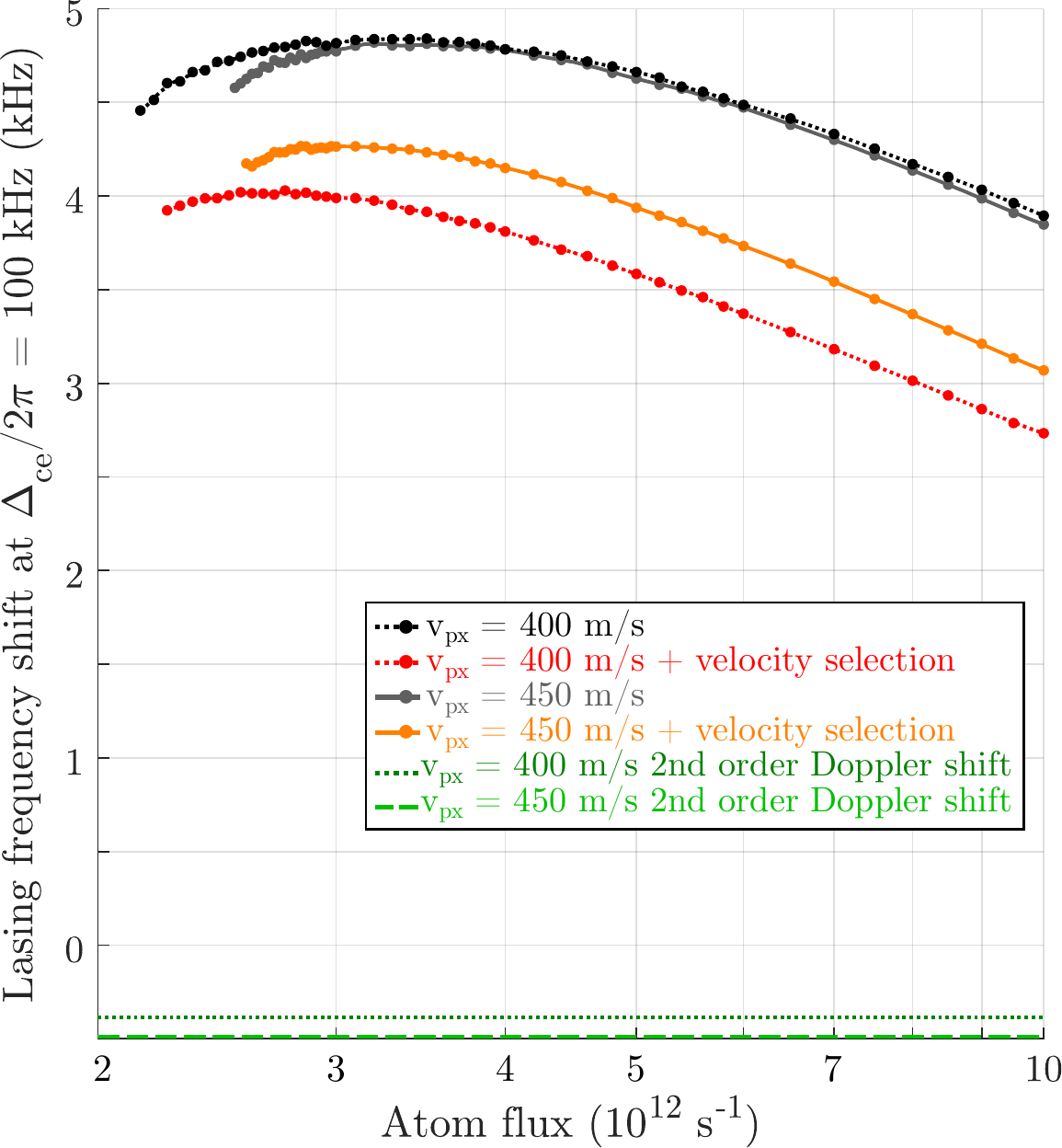}
		\caption{Lasing frequency shift as a function of atomic flux compared for two most probable atomic velocities. The expected frequency shift is shown with and without the velocity selection scheme for a fixed cavity detuning of 100 kHz. The significant 2nd order Doppler shift in the ensemble lasing frequency is illustrated for the two velocities. The cavity pulling coefficients are below 0.06, and we find the velocity selection scheme reduces cavity pulling 10-25 \% in this regime. Lines show a moving mean. \label{fig:hotBeamdetl}}
	\end{figure}

	Due to the 2nd order Doppler shifts, cavity pulling will occur relative to a frequency that is shifted depending on the 2nd order Doppler shifts of the atoms in the ensemble. As the atoms interact differently depending on their velocity, and this interaction also depends on the atom flux, the exact resonance frequency of the ensemble is nontrivial, but it will be close to the most probable 2nd order Doppler shift. We define the pulling coefficient $c_{pull}$ = $(\Delta_L - \Delta_D)/(\Delta_{ce} - \Delta_D) \approx (\Delta_L - \Delta_D)/\Delta_{ce}$, where $\Delta_D$ is the shift in the atom ensemble resonance caused by 2nd order Doppler shifts. The approximation holds for $\Delta_{ce} \gg \Delta_D$. The cavity pulling characteristics are shown in Fig. \ref{fig:hotBeamdetl} for a fixed cavity detuning of 2$\pi \times$ 100 kHz. Here we find cavity pulling coefficients in the range of about 0.03 to 0.06, which are locally independent of the atom flux slightly above threshold. These pulling coefficients represent the behavior for small detunings, but in general the pulling coefficients also vary for changes in detuning that are on the order of the cavity linewidth. We also see that the velocity selection scheme reduces cavity pulling on the order of 15\%, though in a system where the cutoff velocity is considered a free parameter one could reduce this further at higher flux values to reduce cavity pulling further, at the cost of a lower output power. In the simulations with $v_{px}$ = 450 m/s the velocity selection scheme has a smaller impact, as more atoms follow the selection criterium.
	
	For the hot beam system considered here, the Purcell rate is $2 \pi \cross$ 9.4 Hz. As for the cold beam system, the simulations neglect spontaneous emission into the cavity mode, but do not indicate a broadening of the linewidth from other sources.
	
	\section{Conclusion}
	We find that the hot beam approach can realistically produce an output power of hundreds of nW to µW on the $^1$S$_0$-$^3$P$_1$ transition in $^{88}$Sr, and the hot beam system investigated here is capable of suppressing cavity noise by a factor of 20-30. The exact lasing frequency is shifted by approximately 500 Hz due to the 2nd order Doppler shift. On the other hand, this effect is negligible in the cold-atom system. This system is significantly more complex and relies on repumping of atoms within the optical cavity to provide a power on the order of hundreds of pW. However, it could realistically suppress cavity noise by a factor of 50-100. It furthermore has a number of experimental parameters which can be optimized to minimize the sensitivity of the lasing frequency to fluctuations in the atom flux. Both approaches using the $^3$P$_1$ state in Sr to generate superradiant lasing are promising as reference laser candidates that can provide output power levels high enough to be easily detectable, which is advantageous compared to much more narrow clock lines.
	
	\begin{acknowledgments}
		The authors would like to thank Shayne Bennetts and Florian Schreck for useful discussions, and the rest of the iqClock group at UvA for their ideas and collaboration on the hot beam project. This project has received funding from the European Union’s (EU) Horizon 2020 research and innovation programme under grant agreement No 820404 (iqClock project), the USOQS project (17FUN03) under the EMPIR initiative, and the Q-Clocks project under the European Comission's QuantERA initiative. It was additionally supported by a research grant 17558 from VILLUM FONDEN. SAS would like to thank the Independent Research Fund Denmark under project No. 0131-00023B.
	\end{acknowledgments}
	
	\appendix*
	\section{Model equations}
	Within first order mean-field theory the expectation value of the cavity photon number is $\expval{n} \approx \expval{a}^*\expval{a}$, where the expectation value of the lowering operator evolves according to (from now on dropping angle brackets for expectation values):
	
    \begin{equation}
	\begin{aligned}
    \dot{a} = -\left(i \Delta_{cE} + \frac{\kappa}{2}\right) a - i\frac{\eta}{2} e^{i \Delta_{dE} t} - i \sum_{j=1}^{N} g_j \sigma_{ge}^j.
	\end{aligned}
	\end{equation}
	
	The detuning with respect to the unperturbed atomic transition frequency (chosen as the rotating reference frame) is $\Delta_{cE} = \omega_c - \omega_E$, the cavity linewidth is $\kappa$, driving laser frequency $\omega_d$ and intensity parameter $\eta$. The equation for $\dot{a}$ couples to the atomic coherences on the lasing transition. These are modeled in the framework of the SME, where we will continue to use the $\sigma$ operator notation, but note they represent stochastic values, as e.g. $\expval{\rho_{eg}^j}=\expval{\sigma_{ge}^j}$. In this framework $\sigma_{ge}^j$ does not continually decay, but instead $\sigma_{ee}^j$ and $\sigma_{ge}^j$ have a probability within a time interval, $p_j(dt) = \sigma_{ee}^j \gamma_{ge} dt$, to make a quantum jump, collapsing to $\sigma_{gg}^j$=1 and resulting in all other populations and coherences being 0. However in the time intervals where this does not happen, the dynamics evolve according to:
	
	\begin{equation}
	\begin{aligned}
    \dot{\sigma}_{ge}^j = &-i\Delta_{aeE}^j \sigma_{ge}^j + i g_j a \left(\sigma_{ee}^j - \sigma_{gg}^j\right)\\
    &+ i\frac{\chi_{li}^j}{2} \sigma_{ie}^j e^{-ik_{li}r_j+i\Delta_{liI}t}\\
    &+ \sigma_{ge}^j \left[-\gamma_{eg} \left(\frac{1}{2}-\sigma_{ee}^j \right)+\gamma_{eg} \sigma_{ii}^j+\gamma_{xi} \sigma_{xx}^j \right],
	\end{aligned}
	\end{equation}
	
	where $\Delta_{aeE}^j = \omega_{ae}^j - \omega_E$, and $\omega_{ae}^j$ is the shifted atomic transition frequency. As there are no Zeeman shifts or significant light shifts on this transition which are not already contained in the coherent dynamics, $\Delta_{aeE}^j$=0. $\gamma_{eg} = 2\pi \cross $ 7.5 kHz is the exited state decay rate and $\chi_{li}^j$ is the atom's Rabi frequency due to the 689 nm pumping laser. $k_{li} r_j$ denotes the dot product of the laser wavevector with the atomic position vector, and $\Delta_{liI}=0$ is the laser detuning from the Zeeman-shifted atomic transition frequency. The last line is a renormalization factor which corrects the coherent dynamics in the SME framework.\\
	
	For each atom these equations for $\dot{\sigma}_{ge}^j$ are eventually coupled to the equations for the 13 relevant atomic levels, which are internally coupled by the pumping scheme and decay paths. These following subscripts are used to refer to these levels:\\
	$g$ = $^1$S$_0$, $n$ = $^3$P$_0$,\\
	$(i,e,u)$ = $^3$P$_1$ m$_J$ = (-1,0,1),\\
	$(x,y,z)$ = $^3$S$_1$ m$_J$ = (-1,0,1),\\
	$(p,q,r,s,t)$ = $^3$P$_2$ m$_J$ = (-2,-1,0,1,2).\\
	For the states $\ket{g}$, $\ket{e}$, $\ket{i}$ and $\ket{x}$, interactions with the cavity mode and pump lasers are also treated including coherences, giving 5 additional coherence equations:
	
   	\begin{eqnarray}
	\begin{aligned}
    \dot{\sigma}_{gi}^j &= i g_j \sigma_{ei}^j a + i \frac{\chi_{li}^j}{2} \left(\sigma_{ii}^j - \sigma_{gg}^j\right) e^{-ik_{li}r_j+i\Delta_{liI}t}\\
    &- i\frac{\chi_{lx}^j}{2} \sigma_{gx}^j e^{ik_{lx}r_j-i\Delta_{lxX}t} \\
    &+\sigma_{gi}^j \left[\gamma_{eg} \sigma_{ee}^j-\gamma_{ig}\left(\frac{1}{2}-\sigma_{ii}^j\right)+\gamma_{xi} \sigma_{xx}^j\right]
	\end{aligned}
	\end{eqnarray}
   	\begin{eqnarray}
	\begin{aligned}
    \dot{\sigma}_{ix}^j &= i \frac{\chi_{li}^j}{2} \sigma_{gx}^j e^{ik_{li}r_j-i\Delta_{liI}t}\\
    &+ i \frac{\chi_{lx}^j}{2} \left(\sigma_{xx}^j - \sigma_{ii}^j\right) e^{-ik_{lx}r_j+i\Delta_{lxX}t}\\
    &+\sigma_{ix}^j \left[\gamma_{eg} \sigma_{ee}^j-\gamma_{ig}\left(\frac{1}{2}-\sigma_{ii}^j\right)-\gamma_{xi} \left(\frac{1}{2}-\sigma_{xx}^j\right)\right]
	\end{aligned}
	\end{eqnarray}
   	\begin{eqnarray}
	\begin{aligned}
    \dot{\sigma}_{gx}^j &= ig_j \sigma_{ex}^j a + i \frac{\chi_{li}^j}{2} \sigma_{ix}^j e^{-ik_{li}r_j+i\Delta_{liI}t}\\
    &- i\frac{\chi_{lx}^j}{2}\sigma_{gi}^j e^{-ik_{lx}r_j+i\Delta_{lxX}t} \\
    &+\sigma_{gx}^j \left[\gamma_{eg} \sigma_{ee}^j+\gamma_{ig}\sigma_{ii}^j-\gamma_{xi} \left(\frac{1}{2}-\sigma_{xx}^j\right)\right]
	\end{aligned}
	\end{eqnarray}
   	\begin{eqnarray}
	\begin{aligned}
    \dot{\sigma}_{ei}^j &= i \Delta_{aeE}^j \sigma_{ei}^j
    + i g_j \sigma_{gi}^j a^\dagger - i\frac{\chi_{li}^j}{2} \sigma_{eg}^j e^{-ik_{li}r_j+i\Delta_{liI}t}\\
    &- i\frac{\chi_{lx}^j}{2}\sigma_{ex}^j e^{ik_{lx}r_j-i\Delta_{lxX}t} \\
    &+\sigma_{ei}^j \left[-\gamma_{eg} \left(\frac{1}{2}-\sigma_{ee}^j\right)-\gamma_{ig}\left(\frac{1}{2}-\sigma_{ii}^j\right)+\gamma_{xi} \sigma_{xx}^j\right]
	\end{aligned}
	\end{eqnarray}
   	\begin{eqnarray}
	\begin{aligned}
    \dot{\sigma}_{ex}^j &= i \Delta_{aeE}^j \sigma_{ex}^j
    + ig_j \sigma_{gx}^j a^\dagger\\
    &- i\frac{\chi_{lx}^j}{2} \sigma_{ei}^j e^{-ik_{lx}r_j+i\Delta_{lxX}t} \\
    &+\sigma_{ex}^j \left[-\gamma_{eg} \left(\frac{1}{2}-\sigma_{ee}^j\right)+\gamma_{ig}\sigma_{ii}^j-\gamma_{xi} \left(\frac{1}{2}-\sigma_{xx}^j\right)\right].
	\end{aligned}
	\end{eqnarray}

	Here the same notation is used for the 688 nm pump laser as for the 689 nm pump, now with $x$ in place of $i$. The equations for the four related populations are:
	\begin{eqnarray}
	\begin{aligned}
	\dot{\sigma}_{gg}^j &= -ig_j \left(\sigma_{ge}^j a^\dagger - \sigma_{eg}^j a\right)\\
	&- i\frac{\chi_{li}^j}{2} \left(\sigma_{gi}^j e^{ik_{li}r_j-i\Delta_{liI}t} - \sigma_{ig}^j e^{-ik_{li}r_j+i\Delta_{liI}t} \right) \\
    &+\sigma_{gg}^j \left[\gamma_{eg} \sigma_{ee}^j+\gamma_{ig}\sigma_{ii}^j+\gamma_{xi} \sigma_{xx}^j\right] \\
    \dot{\sigma}_{ee}^j &= ig_j \left(\sigma_{ge}^j a^\dagger - \sigma_{eg}^j a\right) \\
    &+\sigma_{ee}^j \left[-\gamma_{eg} \left(1-\sigma_{ee}^j\right)+\gamma_{ig}\sigma_{ii}^j+\gamma_{xi} \sigma_{xx}^j\right] \\
    \dot{\sigma}_{ii}^j &= i\frac{\chi_{li}^j}{2} \left(\sigma_{gi}^j e^{ik_{li}r_j-i\Delta_{liI}t} - \sigma_{ig}^j e^{-ik_{li}r_j+i\Delta_{liI}t} \right) \\
    &- i\frac{\chi_{lx}^j}{2} \left(\sigma_{ix}^j e^{ik_{ix}r_j-i\Delta_{lxX}t} - \sigma_{xi}^j e^{-ik_{lx}r_j+i\Delta_{lxX}t} \right)\\
    &+\sigma_{ii}^j \left[\gamma_{eg} \sigma_{ee}^j -\gamma_{ig}\left(1-\sigma_{ii}^j\right)+\gamma_{xi} \sigma_{xx}^j\right] \\
    \dot{\sigma}_{xx}^j &= i\frac{\chi_{lx}^j}{2} \left(\sigma_{ix}^j e^{ik_{lx}r_j-i\Delta_{lxX}t} - \sigma_{xi}^j e^{-ik_{lx}r_j+i\Delta_{lxX}t} \right) \\
    &+\sigma_{xx}^j \left[\gamma_{eg} \sigma_{ee}^j +\gamma_{ig}\sigma_{ii}^j +\gamma_{xi} \left(1-\sigma_{xx}^j\right)\right].
	\end{aligned}
	\end{eqnarray}
	
	The decays which determine the probability of quantum jumps into and out of these four states are equivalent to the following rate equations:
	\begin{eqnarray}
	\begin{aligned}
	\dot{\sigma}_{gg}^j &= \gamma_{eg} \left( \sigma_{ii}^j + \sigma_{ee}^j + \sigma_{uu}^j \right)\\
    \dot{\sigma}_{ee}^j &= -\gamma_{eg} \sigma_{ee}^j + \frac{\gamma_{xe}}{2} \left( \sigma_{xx}^j + \sigma_{zz}^j \right)\\
    \dot{\sigma}_{ii}^j &= -\gamma_{ig} \sigma_{ii}^j + \frac{\gamma_{xi}}{2} \left(\sigma_{xx}^j + \sigma_{yy}^j\right)\\
    \dot{\sigma}_{xx}^j &= w_{nx}^j \sigma_{nn}^j
    + w_{px}^j \sigma_{pp}^j + w_{rx}^j \sigma_{rr}^j
    - \gamma_x \sigma_{xx}^j.
	\end{aligned}
	\end{eqnarray}
	
	Here $\gamma_x$ with single subscript refers to the total decay rate out of $\ket{x}$. The remaining states are treated using rate equation approximations which are used to implement discrete quantum jumps not only for decays, but also for excitations by the pumping lasers. The rate equation pumping rates are denoted by $w$. As the $^3$S$_1$ populations decay rapidly, we neglect de-excitations by pumping lasers to these levels, and obtain for the remaining two:
   	\begin{eqnarray}
	\begin{aligned}
    \dot{\sigma}_{yy}^j
    &= w_{qy}^j \sigma_{qq}^j+ w_{sy}^j\sigma_{ss}^j
    - \gamma_y \sigma_{yy}^j \\
    \dot{\sigma}_{zz}^j
    &= w_{nz}^j \sigma_{nn}^j
    + w_{uz}^j \sigma_{uu}^j \\
    &+ w_{rz}^j \sigma_{rr}^j + w_{tz} \sigma_{tt}^j
    - \gamma_z \sigma_{zz}^j.
	\end{aligned}
	\end{eqnarray}
	
	This approach is also used for the long-lived states $^3$P$_0$ and $^3$P$_2$, and the remaining m$_J$=1 level of $^3$P$_1$, where we obtain:
	\newpage
   	\begin{eqnarray}
	\begin{aligned}
    \dot{\sigma}_{nn}^j
    = &-w_{nx}^j \sigma_{nn}^j - w_{nz}^j \sigma_{nn}^j
    + \gamma_{xn} \left( \sigma_{xx}^j+ \sigma_{yy}^j + \sigma_{zz}^j \right) \\
    \dot{\sigma}_{pp}^j
    = &-w_{px}^j \sigma_{pp}^j
    + \frac{6}{10} \gamma_{xp} \sigma_{xx}^j \\
    \dot{\sigma}_{qq}^j
    = &-w_{qy}^j \sigma_{qq}^j
    + \frac{3}{10} \gamma_{xp} \left( \sigma_{xx}^j + \sigma_{yy}^j \right) \\
    \dot{\sigma}_{rr}^j
    = &-w_{rx}^j \sigma_{rr}^j - w_{rz}^j \sigma_{rr}^j \\
    &+ \gamma_{xp} \left( \frac{\sigma_{xx}^j}{10} + \frac{4\sigma_{yy}^j}{10} + \frac{\sigma_{zz}^j}{10} \right) \\
    \dot{\sigma}_{ss}^j
    = &-w_{sy}^j \sigma_{ss}^j
    + \frac{3}{10} \gamma_{xp} \left( \sigma_{yy}^j + \sigma_{zz}^j \right) \\
    \dot{\sigma}_{tt}^j
    = &-w_{tz}^j \sigma_{tt}^j
    + \frac{6}{10} \gamma_{xp} \sigma_{zz}^j \\
    \dot{\sigma}_{uu}^j
    = &-w_{uz}^j \sigma_{uu}^j
    - \gamma_{eg} \sigma_{uu}^j + \frac{\gamma_{xu}}{2}\left(\sigma_{yy}^j + \sigma_{zz}^j \right).
	\end{aligned}
	\end{eqnarray}
	
	Finally the filter cavity annihilation operators $f^k$, which are used for calculating the spectrum, evolve according to:
    \begin{eqnarray}
	\begin{aligned}
    \dot{f}_k &= -i \Delta_{fE}^k f_k - iG a,
	\end{aligned}
	\end{eqnarray}
	
	where G is the interaction rate with the main cavity, which can be arbitrary (resulting in a scaling factor for the spectral intensity) when back-action on the main cavity is neglected.
	

\end{document}